\documentclass[twoside]{amsart}
\usepackage{latexsym}
\usepackage{amssymb,amsmath,amsopn}
\usepackage[dvips]{graphicx}   
\usepackage{color,epsfig}      
\usepackage{subcaption}
\usepackage{siunitx}

               {\begin{list}{}{\leftmargin#1\rightmargin#2}\item{}}%
               {\end{list}}

\def\xR{\mathbb{R}}

\def\x0{\mathbf{0}}
\def\xa{\mathbf{a}}
\def\xf{\mathbf{f}}

\def\xn{\mathbf{n}}
\def\xu{\mathbf{u}}
\def\xe{\mathbf{e}}

\def\xC{\mathbf{C}}

\def\xF{\mathbf{F}}
\def\xG{\mathbf{G}}
\def\xI{\mathbf{I}}
\def\xK{\mathbf{K}}

\def\xN{\mathbf{N}}

\def\xxi{\boldsymbol{\xi}}

\DeclareMathOperator{\tr}{tr}

\begin{document}
\title[The Mullins effect in the wrinkling behavior of highly stretched...]{The Mullins effect in the wrinkling behavior of highly stretched thin films}
\author[E. Feh\'er, T. J. Healey  \and A. \'A. Sipos]{Eszter Feh\'er, Timothy J. Healey \and Andr\'as \'Arp\'ad Sipos}

\address{Eszter Feh\'er, MTA-BME Morphodynamics Research Group and Dept. of Mechanics, Materials and Structures, Budapest University of Technology,
M\H uegyetem rakpart 1-3., Budapest, Hungary, 1111}
\email{fehereszter@szt.bme.hu}
\address{Timothy J. Healey, Department of Mathematics, Cornell University, Ithaca, 14850 USA}
\email{tjh10@cornell.edu}
\address{Andr\'as \'A. Sipos, MTA-BME Morphodynamics Research Group and Dept. of Mechanics, Materials and Structures, Budapest University of Technology,
M\H uegyetem rakpart 1-3., Budapest, Hungary, 1111}
\email{siposa@eik.bme.hu}


\begin{abstract}

Recent work demonstrates that finite-deformation nonlinear elasticity is essential in the accurate modeling of wrinkling in highly stretched thin films. Geometrically exact models predict an isola-center bifurcation, indicating that for a bounded interval of aspect ratios only, stable wrinkles appear and then disappear as the macroscopic strain is increased. This phenomenon has been verified in experiments. In addition, recent experiments revealed the following striking phenomenon: For certain aspect ratios for which no wrinkling occurred upon the first loading, wrinkles appeared during the first unloading and again during all subsequent cyclic loading. Our goal here is to present a simple pseudo-elastic model, capturing the stress softening and residual strain observed in the experiments, that accurately predicts wrinkling behavior on the first loading that differs from that under subsequent cyclic loading. In particular for specific aspect ratios, the model correctly predicts the scenario of no wrinkling during first loading with wrinkling occurring during unloading and for all subsequent cyclic loading.
\end{abstract}
\maketitle

\section{Introduction}
\label{sec:intro}
The need for finite-deformation nonlinear elasticity in the accurate modeling of wrinkling phenomena in highly axially stretched thin elastomer sheets was recently demonstrated, cf. \cite{Healey, Li, Sipos}. The 2D membrane model is that of geometrically exact nonlinear elasticity, while the fine thickness of such sheets manifests itself in an extremely small bending stiffness. This is in contrast to the well-known Fl\"oppl-von K\'arm\'an (FvK) model \cite{Karman1}, employing linear infinitesimal elasticity in the membrane part that also incorporates a nonlinear term in the gradient of the out-of-plane displacement. The FvK model has a long and successful track record in the prediction of buckling and initial post-buckling behavior of classical plates and shells.

\begin{figure}[!ht]
	\begin{center}
	\includegraphics[width=0.85\textwidth]{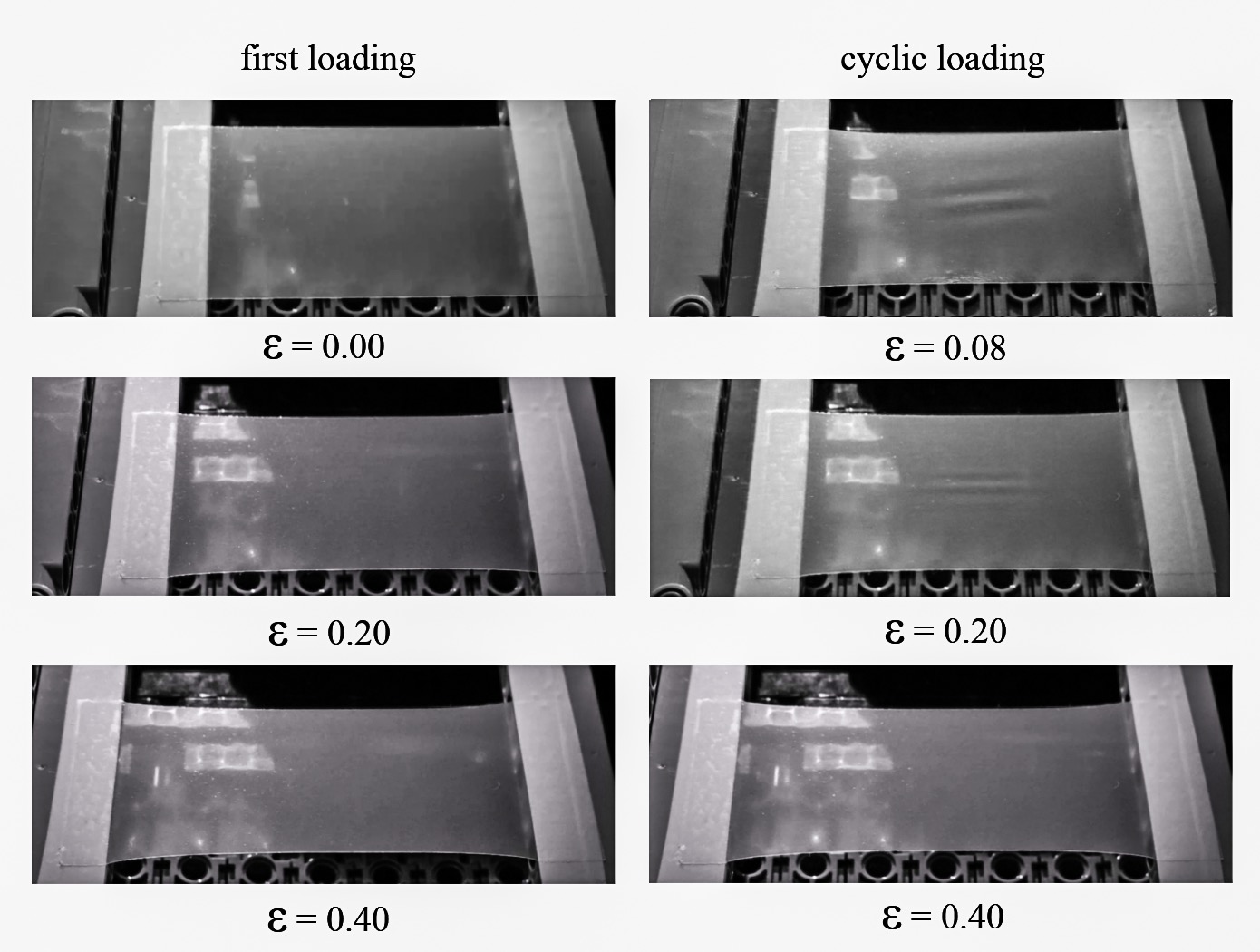}
	\caption{Inelastic behavior of stretched, thin films: a film remains flat during the first loading (left panel) but wrinkles keep appearing and disappearing during the cyclic loading afterwards (right panel). The residual strain is about $\varepsilon=0.08$, observe, that the sheet is still wrinkled at the residual strain (top right figure).} \label{fig:motiv}
	\end{center}
\end{figure} 

Of course the wrinkling of a flat sheet is also a buckling phenomenon. Nonetheless, as shown in \cite{Healey,Li}, the FvK model leads to qualitative errors in the prediction of wrinkles for that particular class of problems. A bifurcation analysis in the macroscopic (applied) strain indicates that, for an apparently semi-bounded interval of aspect-ratio values, the FvK theory predicts a super-critical pitch-fork bifurcation indicating the initiation of wrinkles. In particular, it fails to predict the disappearance of wrinkles as the macroscopic strain is increased beyond initiation. In contrast, the geometrically exact models of \cite{Healey,Li} predict an isola-center bifurcation, indicating that for a bounded interval of aspect ratios only, stable wrinkles appear and then disappear as the macroscopic strain is increased. To summarize, the FvK model not only predicts the wrong phenomena (wrinkles do not disappear), but it also predicts the initiation of wrinkles for a vast range of aspect ratios not exhibiting wrinkling. 

\begin{figure}[!ht]
	\begin{center}
	\includegraphics[width=0.45\textwidth]{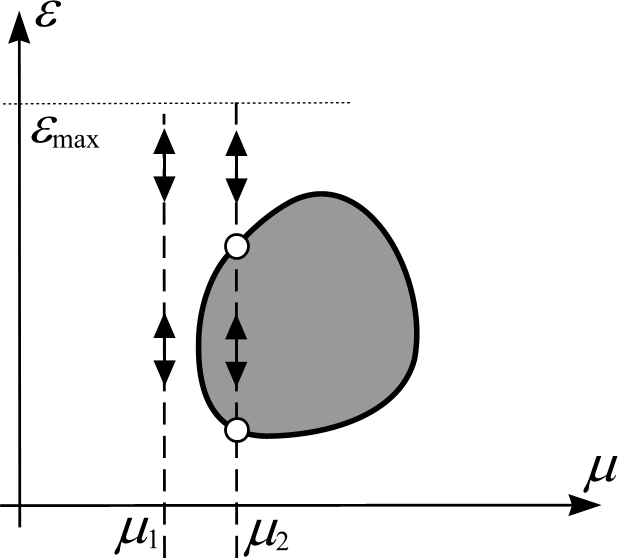}
	\caption{Stability boundary for a purely elastic sheet.}	
	\label{fig:explanation}
	\end{center}
\end{figure} 

In recent experimental work \cite{Feher,Sipos}, the appearance and subsequent disappearance of wrinkles for a bounded range of aspect ratios was unambiguously verified for the problem addressed in \cite{Healey,Li}. This underscores the importance of a geometrically exact mathematical model. However, the experiments reported in \cite{Sipos} reveal inelastic behavior corresponding to a permanent change in length and shape of sheets upon first loading with little change upon subsequent loadings. We also point out that the recent work \cite{Zhu} also calls for an understanding of the role of inelasticity in wrinkling behavior.

Here we report on further experimental results for this problem, carried out on finely thin, rectangular elastomer (polyurethane) sheets. The specimens were subjected to cyclical “hard” loading and unloading. To fix terminology, the first macroscopic stretch of the virgin sheet is called the \emph{first loading}. The first unloading and all subsequent loadings and unloadings are referred to collectively as \emph{cyclic loading}. Among other observations, our work here is motivated by the following striking experimental observation: For certain aspect ratios for which no wrinkling occurred upon the first loading, wrinkles then appeared during unloading and again during all subsequent cyclic loading. (See Figure \ref{fig:motiv}.) Moreover, residual strain and significant stress softening was observed after the first unloading, while all subsequent load cycles exhibited elastic behavior, as pointed out earlier in \cite{Sipos}. Wrinkling aside, the latter is consistent with the so-called Mullins effect in elastomers, e.g. \cite{Dorfman,Ogden}. 

We can give a heuristic explanation of the observed phenomenon, based upon our earlier results for purely elastic models. A typical stability boundary of the type obtained in \cite{Li} is a simple closed curve in the plane of aspect ratio, denoted $\mu=L/2W$, vs. macroscopic strain, denoted $\varepsilon=\Delta L/L$. Here $L$ and $W$ are the length and width, respectively, of the rectangular sheet, while $\Delta L$ is the change in length under externally imposed, longitudinal stretching. In Figure 2 we depict such a computed curve for the specimens studied in this work, employing a finite-elasticity model from \cite{Li}. The intersection of a vertical line, corresponding to a fixed aspect ratio, with the closed curve gives the locations where stable wrinkles first emerge and then disappear as $\varepsilon$ is steadily increased. In particular, no wrinkling occurs for aspect ratios below or above the closed curve. With that in mind, suppose that a given sheet has aspect ratio $\mu_1$ that is just below but sufficiently close to the stability boundary. Now assume that the first loading produces a permanent change of length such that the aspect ratio now increases to $\mu_2$ as shown in Figure \ref{fig:explanation}. Then wrinkling occurs upon unloading.

Of course, the qualitative explanation above is a gross oversimplification. The purely elastic model neither captures the developed orthotropy observed in \cite{Sipos}, nor does it account for the change in the stability boundary due to damage. In particular, the entire specimen goes slack at a small positive value of the macroscopic strain upon unloading, corresponding to the residual strain in the specimen. Moreover, it was observed that the unloaded specimens were wrinkled at the residual strain level. This is illustrated in the top right panel of Figure \ref{fig:motiv}; the residual strain is about $\varepsilon=0.08$. Our goal here is to present a simple pseudo-elastic model, accounting for the longitudinal stress softening and residual strain observed in the experiments, that accurately captures the correct wrinkling behavior just described. In particular for specific aspect ratios, the model correctly predicts the scenario of no wrinkling during first loading with wrinkles developing upon unloading and during all subsequent cyclic loading.  

The outline of the paper is as follows. In Section 2 we present our model incorporating a single state variable to capture the Mullins effect as in \cite{Dorfman} for the 2D Mooney-Rivlin elasticity model with small bending employed in \cite{Li}. The dominantly uni-axial Mullins effect is accounted for solely in the membrane part of the model, while in the presence of small elastic bending. In Section 3 we present the pertinent experimental data, some of which are used to first “tune” the mathematical model in Section 4. We then compare the experimental results for the appearance and disappearance of wrinkles for a large set of aspect ratios with our numerical predictions, demonstrating the accuracy of the mathematical model. We present some concluding remarks in Section 5.

\section{Model development}
\label{sec:model_development}

\begin{figure}[!ht]
	\begin{center}
	\includegraphics[width=0.70\textwidth]{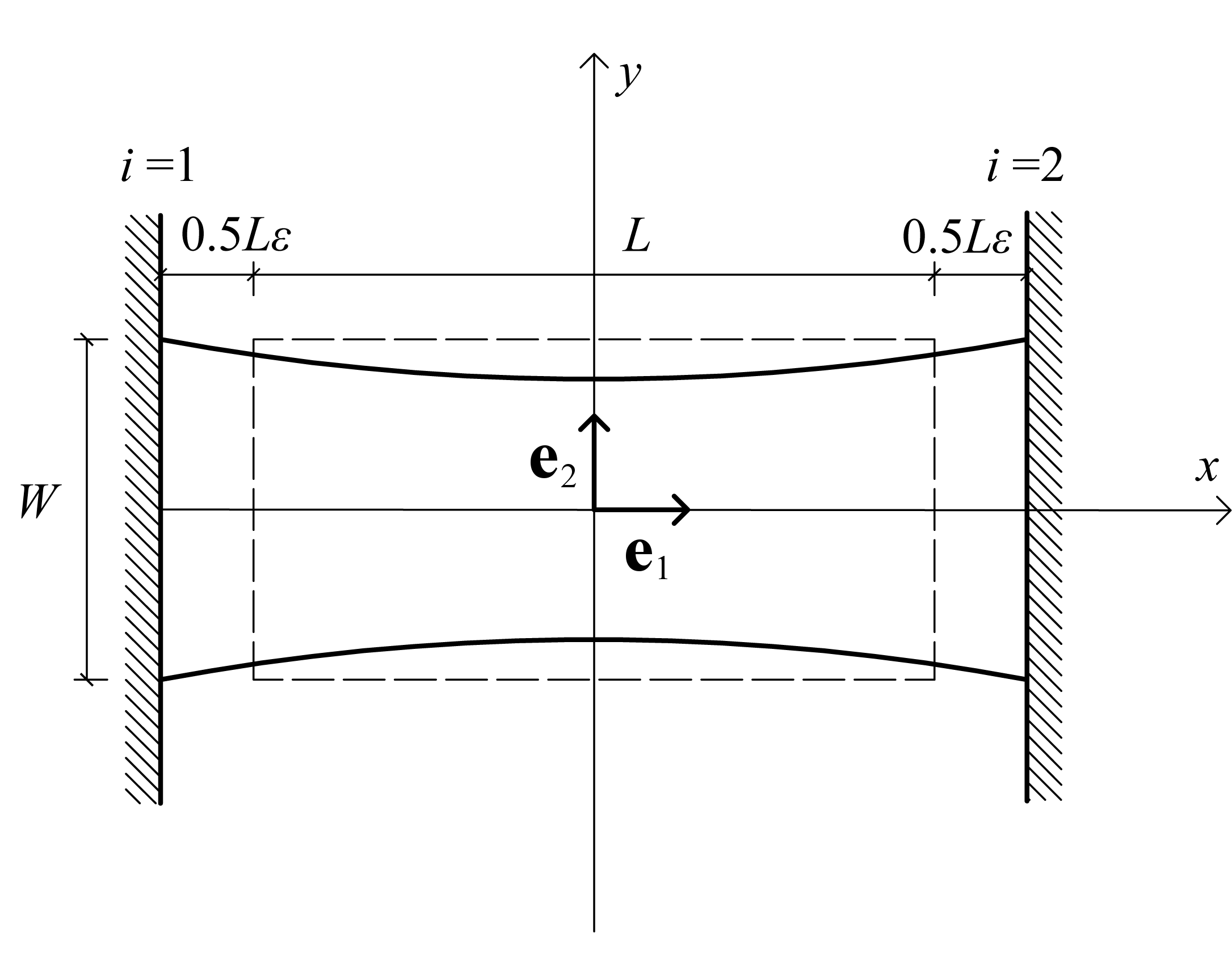}
	\caption{A stretched, rectangular sheet clamped along its opposite edges}
	\label{fig:config}
	\end{center}
\end{figure}

We let $\Omega\subset\xR^2$ denote a rectangular domain of length of $L$ and width of $W$, which we choose as a stress-free reference configuration (Figure \ref{fig:config}.). The displacement field $\xu:\Omega\rightarrow \xR^3$ is denoted by
\begin{equation}
	\xu:=\begin{bmatrix}
		u(x,y)\\
		v(x,y)\\
		w(x,y)\\
	\end{bmatrix},
\end{equation}

\noindent and for convenience, the in-plane displacement field is denoted by $\hat{\xu}:\Omega\rightarrow \xR^2\times\{0\}$, i.e. $\hat{\xu}:=[u(x,y), v(x,y),0]^T$.  We henceforth employ the notation $\tilde{\xR}^2:=\xR^2\times\{0\}\subset\xR^3$. The sheet is subjected to hard-loading: Displacements are prescribed on opposite ends via $\xu(\pm L/2,y)=[\pm\varepsilon L/2,0,0]^T$ for $-W/2\leq y \leq W/2$, while the top and bottom at $y=\pm W/2$, respectively, are unconstrained. We call $\varepsilon > 0$ the \emph{macroscopic strain}.

The stored energy of the thin hyperelastic sheet is taken as the sum of membrane and bending energies expressed as functionals of the displacement field $\xu$ \cite{Howell}:
\begin{equation}
I(\xu)=\int_{\Omega}\Psi d\Omega=\int_{\Omega}\Psi_m d\Omega+\int_{\Omega}\Psi_b d\Omega,
\end{equation}

\noindent where the stored energy densities of $\Psi_m$ and $\Psi_b$ denote the constitutive laws for the membrane and the bending behavior, respectively. Following \cite{Li}, we propose an objective stored-energy density, $\Psi_m$ that depends on the right Cauchy-Green strain $\xC=\xF^T\xF$, where the deformation gradient $\xF$ is given by
\begin{equation}
\label{model:F}
\xF=\xI+\nabla \hat{\xu}+\xe_3\otimes\nabla w.
\end{equation}

\noindent Here all gradients are taken with respect to the reference domain $\Omega\subset\xR^2$, $\xI\in L(\xR^2,\tilde{\xR}^2)$ is isomorphic to the identity on $\xR^2$, $\nabla\hat{\xu}(x,y)\in L(\xR^2,\tilde{\xR}^2)$, $\xe_3=[0,0,1]^T$, and ``$\otimes$" denotes the tensor product. Hence, $\xF\in L(\xR^2,\xR^3)$ and $\xC\in L(\xR^2)$.

Our goal is to introduce a pseudo-elastic model that accounts for the stress-softening / residual strain and especially the wrinkling behavior observed in the experiments. In spite of the dominantly uniaxial nature of the experiment, it has been observed that the effective modulus of elasticity drops significantly in both the longitudinal and lateral directions after loading, cf. \cite{Sipos}. Of course that drop is larger in the highly stretched direction. Accordingly, we propose a model incorporating a single state variable acting anisotropically in the $\xe_1$ and $\xe_2$ directions. In particular, the “damage” or state variable acts with a larger magnitude in the highly stretched direction. Correctly tuned, this approach is sufficient to exhibit the observed residual strain and wrinkling behavior. Following the lead of \cite{Dorfman,Ogden}, our targeted model has the form:
\begin{equation}
\label{model:ED}
\Psi=\Psi_m(\xC,\eta)+\Phi(\eta)+\Psi_b. 
\end{equation}

\noindent where $\eta$ is a state variable with $\eta\leq 1$ and $\Phi$ is the so-called  \emph{dissipation function}. Here $\eta=1$ is associated with undamaged material. Predictions based on the hyperelastic models from \cite{Li} were compared to the experimental data on the first loading. For $\Psi_m(\xC,\eta)$ we choose the incompressible Mooney-Rivlin model employed in \cite{Li}, which provides acceptable agreement with our experimental data. Accordingly we propose the following form for the membrane energy density:
\begin{align}
\label{model:mED}
	\nonumber\Psi_m(\xC,\eta):=	&h\alpha\left[\left((1+d)\eta-d\right)(C_{11}-1)+\eta(C_{22}-1)+\frac{1}{\det{\xC}}-1\right]+\\
	&h\beta\eta\left[\frac{\tr\xC}{\det{\xC}}+\det{\xC}-3\right],
\end{align}
	
\noindent where $\alpha$ and $\beta$ are material constants and $d>0$ is a fixed scalar parameter used to tune the anisotropic damage ratio. (Observe from (\ref{model:mED}) that $d=0$ corresponds to isotropic damage.) A completely damaged model corresponds to $\eta=d/(1+d)$. Of course in (\ref{model:mED}) we are associating the components $C_{11}$ and $C_{22}$ with the two principal directions -- a good approximation given the dominantly uniaxial nature of our problem. As such, we show in the appendix that our model is a special case of the general class of models presented in \cite{Dorfman}.

When $\eta\equiv 1$ observe that (\ref{model:mED}) reduces to the thin-membrane approximation of a Mooney-Rivlin elastic solid. Starting from a thin 3D layer, local incompressibility implies that the through-thickness stretch is the reciprocal of the product of the other two stretches, which are assumed to be functions of $x$ and $y$ only. The pressure is subsequently eliminated via stress-free conditions on the two lateral surfaces. This particular model was confirmed by Treolar in bi-axial experiments on thin sheets \cite{Treloar}. We also refer to \cite{MullerStrehlow} for the details and a discussion about the experimental results of Treolar. It’s also worth noting that (\ref{model:mED}) possesses the correct behavior that $\Psi_m \nearrow \infty$ as $\det{\xC} \searrow 0$, and as a purely 2D model in $\xR^2$, it is polyconvex for $\eta = 1$ and $\alpha > 3\beta/4$, i.e., it is a proper, physically reasonable model for 2D nonlinear elasticity.

As for the bending energy, we make the following simple choice:							
\begin{equation}
	\label{model:bED1}
		\Psi_b(\xK)=\frac{Eh^3}{24(1-\nu^2)}\left[\nu(\tr{\xK})^2+(1-\nu)\xK\cdot\xK\right],
\end{equation}

\noindent where $\xK:=-\nabla^2 w$ with $\nabla^2(.):=\nabla\circ\nabla(.)$ denoting the second gradient and $\xK\cdot\xK=\tr{\xK^2}$. In (\ref{model:bED1}) $E$ is the modulus of elasticity and $\nu$ is Poisson's ratio; to keep consistency with the incompressible membrane model, we take $E=6(\alpha+\beta)$ and $\nu=0.5$. Our justification for (\ref{model:bED1}) is the same as that given in \cite{Li}: Taking the gradient of (\ref{model:F}), we arrive at the $3^{\textrm{rd}}$-order tensor-valued field
\begin{equation}
\label{model:bED2}
\xG:=\nabla\xF=\nabla^2\hat{\xu}+\xe_3\otimes\nabla^2 w,
\end{equation}

\noindent where $\nabla^2 w(x,y)\in S^{2\times 2}$ (symmetric $2\times 2$ matrices), $\nabla^2\hat{\xu}(x,y)\in \tilde{\xR}^2\otimes S^{2\times 2}$, and hence $\xG \in \xR^3\otimes S^{2\times 2}$. Letting $\xn(x,y)$ denote a unit normal field on the deformed surface $\xf(\Omega)$, it’s not hard to show that the relative curvature tensor (pull-back of the 2$^{\textrm{nd}}$ fundamental form) is given by $-\xn\cdot\xG$, the product of which is defined as follows: For any $\xG\in \xR^3\otimes S^{2\times 2}$, we have $\sum_{i=1}^3G_{i\alpha\beta}\xe_i=\sum_{i=1}^3G_{i\beta\alpha}\xe_i,\textrm{ } \alpha,\beta=1,2$. Then for any $\xa\in\xR^3, \xa\cdot\xG\in S^{2\times 2}$ has components $\sum_{i=1}^3a_iG_{i\alpha\beta}=\sum_{i=1}^3a_iG_{i\beta\alpha}, \textrm{ } \alpha,\beta=1,2$. As pointed out in \cite{Li}, the wavelength of a typical wrinkle in this setting is two orders of magnitude greater than the maximum wrinkling amplitude, and the latter is of the same order of magnitude as the fine thickness of the sheet. Hence, a valid approximation is $\xn\cong\xe_3$ in which case $-\xn\cdot\xG\cong-\xe_3\cdot\xG=-\nabla^2 w$. Of course the energy density (\ref{model:bED1}) is the usual bending energy for a linear isotropic thin plate – a consistent quadratic approximation for a more general density depending on the relative curvature. Asymptotic versions of the latter have been derived, e.g., in \cite{Hilgers1,Hilgers2}. But perhaps a more accurate model here would account for the Mullins effect in bending, say, by incorporating the state variable into the bending energy. Another possibility would be to use (\ref{model:bED1}) on the virgin loading, and then incorporate (fixed) elastic orthotropy for all subsequent cyclic loadings. However, given that (\ref{model:bED1}) is $\mathcal{O}(h^3)$ in contrast to the $\mathcal{O}(h)$ dependence of (\ref{model:mED}), none of these refinements are important here in this setting; $h=$\SI{32}{\micro m} and $L=50$mm in our experiments. In fact the membrane energy (\ref{model:mED}) is the crucial part of our model. 

Finally, we need to define the last term in equation (\ref{model:ED}). Following \cite{Dorfman}, the so-called dissipation function $\Phi(\eta)$ is defined implicitly via
\begin{equation}
\label{DissFun}
\frac{\partial\Psi}{\partial\eta}=\frac{\partial\Psi_m(\xC,\eta)}{\partial\eta}+\frac{\text{d}\Phi(\eta)}{\text{d}\eta}=0,
\end{equation}

\noindent and we require $\eta=1$ at the point where the first unloading is initiated. In addition, we need to satisfy $\Phi(1)=0$ along the primary loading path and $\Phi''(\eta)<0$ to any admissible value of $\eta$ \cite{Dorfman}. Let $c_1>0$ and $c_2>0$ be (fixed) material parameters. The state variable field 
\begin{equation}
\label{def:eta1}
\eta:=1-c_1\tanh\left(c_2(W_{\max}-W_i)\right),
\end{equation}

\noindent corresponds to a $\Phi(\eta)$ function, that fulfills all the requirements above. Here, based on equation (\ref{DissFun}) we have
	\begin{eqnarray}
	\label{def:eta1_2}
		W_i:=-\frac{\text{d}\Phi(\eta)}{\text{d}\eta}=\frac{\partial\Psi_m(\xC,\eta)}{\partial\eta}, \\
		\label{def:eta1_3}
		W_{\max}:=-\left.\frac{\text{d}\Phi(\eta)}{\text{d}\eta}\right|_{\varepsilon=\varepsilon_{\max}}=\left.\frac{\partial\Psi_m(\xC,\eta)}{\partial\eta}\right|_{\varepsilon=\varepsilon_{\max}}. 
\end{eqnarray}

\noindent Here $\varepsilon_{\max}$ stands for the maximum applied strain for the loading history. Note that this formalism renders $\eta\equiv 1$ along the primary loading path, as required, since $W_i=W_{\max}$ along the primary loading path.

In the absence of external loading, based on equations (\ref{model:mED})-(\ref{DissFun}), the potential energy of the system is summarized as
\begin{equation}
	I(\xu,\eta)=\int_{\Omega}\Psi(\xC, \xK, \eta) \textnormal{d}\Omega=\int_{\Omega}\Psi_m(\xC,\eta)\textnormal{d}\Omega+\int_{\Omega}\Psi_b(\xK) \textnormal{d}\Omega+\int_{\Omega}\Phi(\eta) \textnormal{d}\Omega.
\end{equation}

Denoting the (second Piola-Kirchhoff) in-plane stress via
\begin{equation}
\xN:=2\frac{\partial\Psi_m}{\partial \xC}(\xC),
\end{equation}

\noindent the first variation of the stored energy respect to the $\xu$ displacement field delivers the weak form of the Euler-Lagrange (equilibrium) equations 
\begin{eqnarray}
\label{eq:governing1}
\int_\Omega\left(\xI+\nabla\hat{\xu}\right)\xN(\xC)\cdot\nabla\xxi\textnormal{d}\Omega=0,\\
\label{eq:governing2}
\int_\Omega\left\{(\alpha+\beta)h^3(\Delta w\Delta\zeta+\nabla^2 w\cdot\nabla^2\zeta)+3\left(\xN\nabla w\right)\cdot\nabla\zeta\right\}\textnormal{d}\Omega=0,
\end{eqnarray}

\noindent for all admissible variations $(\xxi,\zeta)$ of $\xu=(\hat{\xu},w)$, where $\Delta(.)$ denotes the Laplacian on $\xR^2$. Observe that eq. (\ref{DissFun}) renders the variation of the energy respect to the $\eta$ field identically zero, while the evolution of the $\eta$ field is defined via eqs. (\ref{def:eta1})-(\ref{def:eta1_3}).

\section{Experimental results}
\label{Sec:exp}

Based upon elastic model predictions predictions, we expect either two critical values of the macroscopic strain, at which wrinkles appear and disappear, or no critical values -- depending upon aspect ratio \cite{Li}. In the first case, the critical values are denoted by $0<\varepsilon_1<\varepsilon_2$, respectively. Two series of experiments were carried out on \SI{32}{\micro m} thick polyurethane sheets:
\begin{enumerate}
	\item Traditional displacement controlled pull-tests to obtain a stress-strain diagram for the material.
	\item A series of cyclic loading of the specimen with different aspect ratios, where $\varepsilon_2$ was observed visually.
\end{enumerate}

\noindent The maximum applied macroscopic strain applied to all specimens was $\varepsilon_{\text{max}}=0.66$ throughout both test series.

\subsection{Force-displacement measurements}
\begin{figure}[!ht]
\begin{center}
			\includegraphics[width=0.60\textwidth]{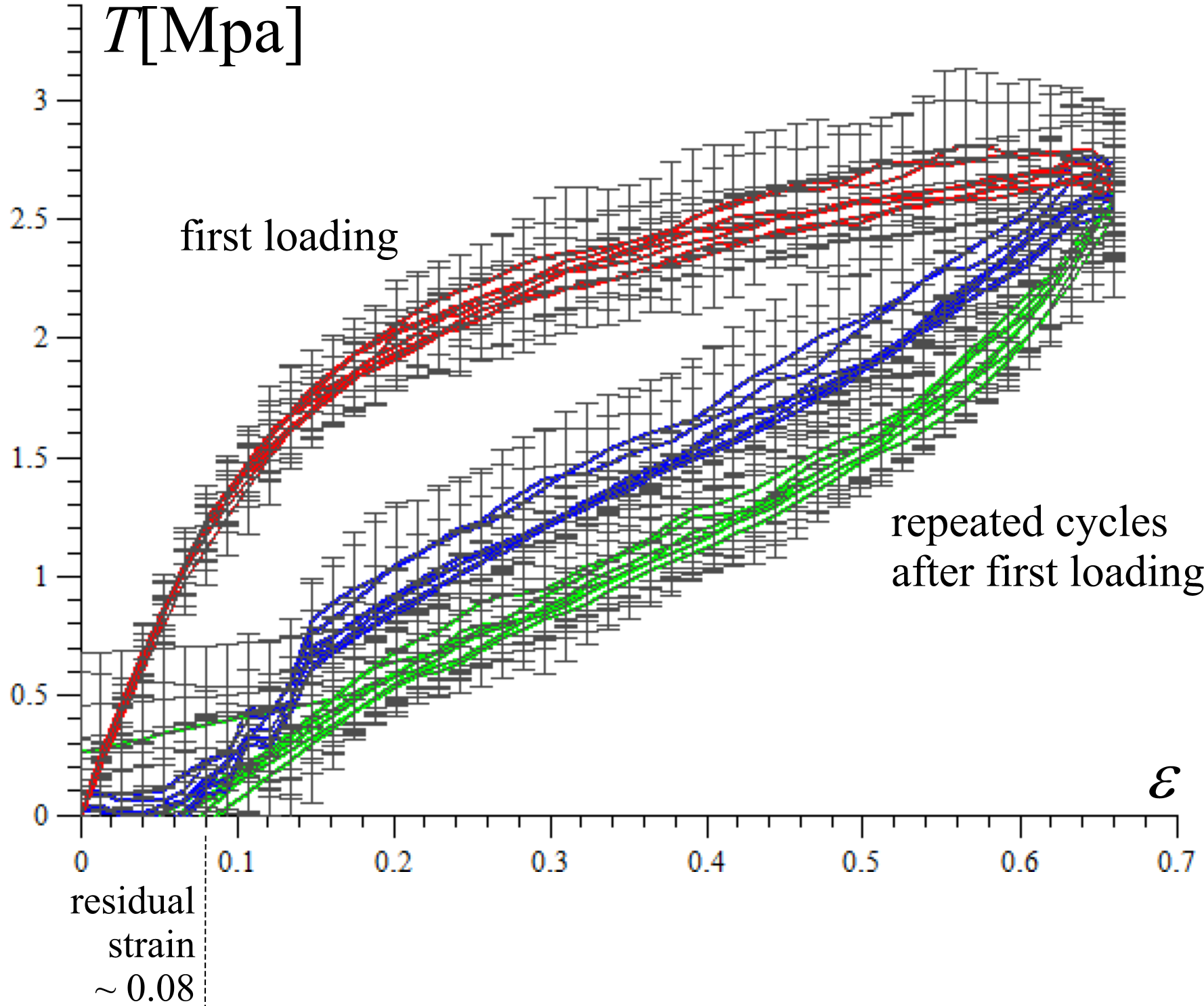}
			\caption{Applied strain ($\varepsilon$) vs. engineering stress ($T$). Red denotes the trend of average stress during the first loading. Blue and green depict the measured averages during cyclic loading in a such way that blue corresponds to loading and green unloading. Data sets measured on sheets with $W\in\{18, 20, 25, 30, 35, 38, 40\}$mm are superimposed.} \label{fig:exp0}
\end{center}
\end{figure} 

Force-displacement measurements were carried out using a Zwick Z150 electromechanical tensile and compression testing machine. The clamped end of the sheet was moved under displacement control at a constant strain rate and the stress-strain diagram was produced. We found no significant dependence of the measured constitutive law on speed or aspect ratio. This observation strongly supports our choice for a pseudoelastic model. The measurements were carried out on $L=50$ mm-long sheets at a speed of $v=120$mm/min. Altogether 28 series of cyclic measurements were performed with 5 loading cycles on each specimen. The results of the first loading and the first cycle are depicted in Figure \ref{fig:exp0}. The observed stress-softening was significant, and about 8\% residual strain was observed. Due to the simplicity of our model, these measurements are sufficient to fit all of the material parameters except the anisotropic constant $d$, cf. (\ref{model:mED}).

\subsection{Critical stretch measurements}
In these series of measurements we determined $\varepsilon_2$, the point of wrinkle disappearance under increased loading. Here the aspect ratio is defined as
\begin{equation}
\mu:=\frac{L}{2W}.
\end{equation}

\begin{figure}
    \centering
    \begin{subfigure}[b]{0.49\textwidth}
			\includegraphics[width=0.99\textwidth]{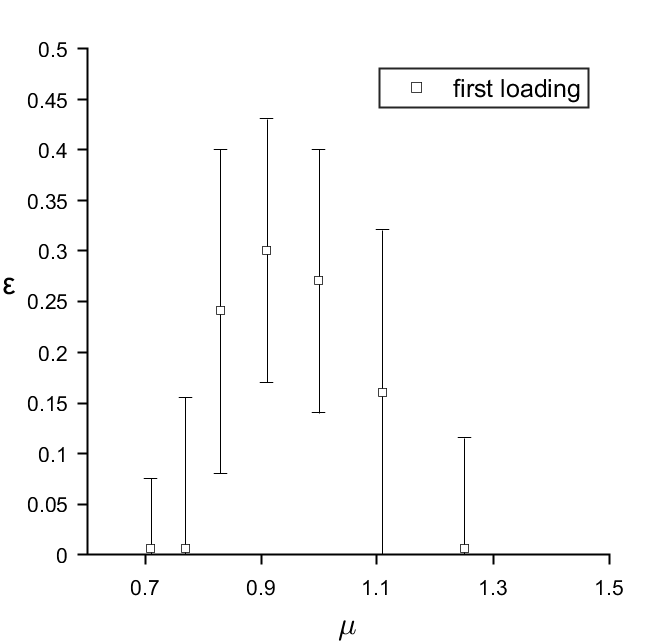}
			\caption{first loading}
    \end{subfigure}
    \begin{subfigure}[b]{0.49\textwidth}
			\includegraphics[width=0.99\textwidth]{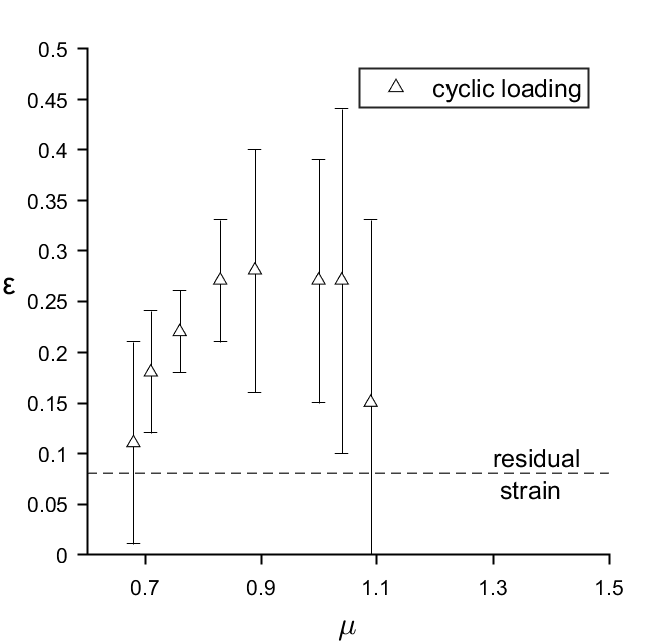}
			\caption{unloading and consecutive loading cycles}
    \end{subfigure}
    \caption{Measured values of disappearance (on unloading) $\varepsilon_2$ vs. $\mu$ with error bars.}\label{fig:exp1}
\end{figure}

\noindent Note that $L$ denotes the length of the virgin sheet. Hence $\mu$ refers to the initial aspect ratio regardless of the modification of the stress-free length. Rectangular sheets with different widths and a fixed length of $L=50$mm were attached to an auxiliary device that enabled repeated pull tests in the horizontal position. Ten to twenty specimens for each width were measured. The collected data, including the standard deviations of the measured values of $\varepsilon_2$, are given in Table \ref{Table:01}. Figure \ref{fig:exp1} depicts the measured values of $\varepsilon_2$ for the first loading (part a) and the cyclic loading (part b), respectively, i.e., $\varepsilon_2$ vs. $\mu$ with error bars. As discussed in the Introduction, the value of $\varepsilon_1$ is only relevant on the first loading; these values could not be measured accurately via visual observation, cf. item 2 before Section 3.1.

\begin{table}[!ht]
\centering
\begin{tabular}{c c | c c | c c}
			\hline
			W    & $\mu$ &\multicolumn{2}{c}{first loading} & \multicolumn{2}{c}{cyclic loading}\\
			mm &  & $\varepsilon_2$ & std & $\varepsilon_2$ & std\\
			\hline
			10 & 2.50 & 0.00 & 0.00 & n.a. & n.a. \\
			15 & 1.67 & 0.00 & 0.05 & n.a. & n.a. \\
			18 & 1.39 & n.a. & n.a. & 0.00 & 0.00 \\
			20 & 1.25 & 0.00 & 0.11 & 0.02 & 0.03 \\
			21 & 1.19 & n.a. & n.a. & 0.05 & 0.07 \\
			23 & 1.09 & 0.16 & 0.16 & 0.15 & 0.18 \\
			25 & 1.00 & 0.27 & 0.13 & 0.27 & 0.12 \\
			28 & 0.91 & 0.30 & 0.13 & 0.28 & 0.12 \\ 
			30 & 0.83 & 0.24 & 0.15 & 0.27 & 0.06 \\
			33 & 0.76 & 0.00 & 0.16 & 0.22 & 0.04 \\  
			35 & 0.71 & 0.00 & 0.07 & 0.18 & 0.06 \\
			37 & 0.68 & n.a. & n.a. & 0.11 & 0.10 \\
			40 & 0.63 & 0.00 & 0.00 & 0.00 & 0.00 \\
			\hline
		\end{tabular}
	
	\caption{Measured values of $\varepsilon_2$ with standard deviations for the first and cyclic loading measured for $L=50$mm rectangular specimen. See Figure \ref{fig:exp1}.}
	\label{Table:01}
\end{table}

\section{Model vs. experiments}
As mentioned earlier, our problem is essentially uniaxial, and values of the material parameters $\alpha, \beta, c_1$ and $c_2$ can be inferred from our uniaxial-test data. In light of incompressibility, this is facilitated by first introducing $\lambda:=1+\varepsilon$, where $\lambda=\lambda_1$ is the principal stretch assumed to be aligned with the $x_1$ axis, and then setting
\begin{equation}
\label{eq:C}
\xC:\cong\begin{bmatrix}
		\lambda_1^2 & 0\\
		0 & \lambda_2^{2}
\end{bmatrix}
\cong
\begin{bmatrix}
		\lambda^2 & 0\\
		0 & \lambda^{-1}
\end{bmatrix}.
\end{equation}

\noindent With these simplifications, the engineering stress (force per unit reference area) in the $x$ direction, denoted by $T_0$ is found from (\ref{model:mED}) to be
	\begin{eqnarray}
	\label{eq:T0}
		T_0(\lambda)=\frac{2\alpha(d+\eta-d\eta)\lambda^4\eta-\alpha(1+\eta)\lambda+2\beta\lambda^3\eta-2\beta\eta}{\lambda^3}.
	\end{eqnarray}

\begin{figure}[!ht]
\begin{center}
			\includegraphics[width=0.70\textwidth]{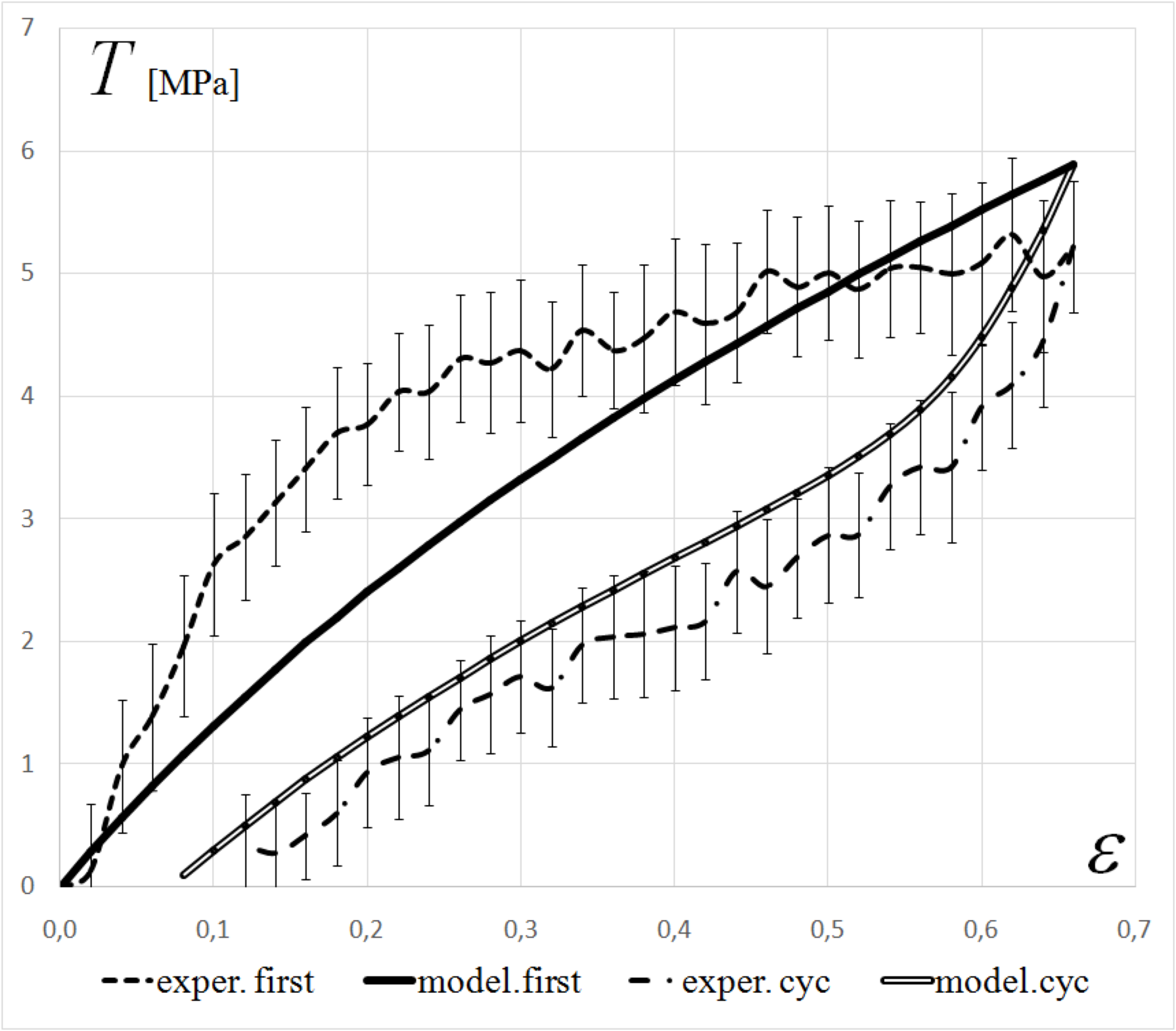}
			\caption{Applied strain ($\varepsilon$) vs. engineering stress ($T$). Dashed lines and error bars correspond to the measured data; solid lines depict model prediction at $\alpha=2.00, \beta=0.45, c_1=0.12, c_2=0.80$ and $d=1.25$.} \label{fig:result0}
\end{center}
\end{figure}

\noindent Because $\eta\equiv 1$ along the first loading, $c_1$ and $c_2$ are inactive. Hence, $\alpha$ and $\beta$ are fitted for the stress-strain curve for the primary loading. The constants $c_1$ and $c_2$ are determined based on the cyclic loading. The best match (with minimal least-squares error between measured and computed data) was achieved at $\alpha=2.00, \beta=0.45, c_1=0.12, c_2=0.80$ (Figure \ref{fig:result0}). The obtained values for $\alpha$ and $\beta$ are in good agreement with earlier results on polyurethane sheets \cite{Spathis, Destrade}. Observe that the model recovers the measured $8\%$ residual strain, and the state variable field drops to approximately $\eta=0.88$ over the most of the sheet as the stress is released upon unloading. The anisotropic ratio of damage was obtained by computing the bifurcation points $\varepsilon_2$ for different aspect ratios: The best match was achieved at $d=1.25$. 

With the parameter values for the constitutive law in hand, we carried out numerical computations in the finite-element-based solver FEniCS 1.6.0 \cite{fenics}. Recall, that $\xxi$ and $\zeta$ denote the admissible variations of $\hat{\xu}$ and $w$, respectively. Let ${(.)_\Delta}$ refer to finite dimensional approximation of a field. The weak form of the governing equations now reads
	
\begin{eqnarray}
\label{eq:weak}
		\nonumber F(\hat{\xu}_\Delta,w_\Delta)=\int_{\Omega}\left[\left(\xI+\nabla\hat{\xu}_\Delta\right)\xN_\Delta\cdot \nabla\xxi_\Delta\right]\textnormal{d}\Omega+&\\
		\int_{\Omega}\left[(\alpha+\beta)(\Delta w_\Delta \Delta\zeta_\Delta+\nabla^2w_\Delta\cdot \nabla^2\zeta_\Delta)+3\xN_\Delta\nabla w_\Delta\cdot\nabla\zeta_\Delta\right]\textnormal{d}\Omega=&0.
\end{eqnarray}

We employ first-order and second-order polynomial approximation for $\hat{\xu}_\Delta$, $\hat{\xi}_\Delta$ and $w_\Delta$, $\zeta_\Delta$, respectively. The domain $\Omega$ is discretized with a uniform, triangular mesh consisting approximately 10000 finite elements with 30000 DOF. Since eq. (\ref{eq:weak}) depends on the second derivatives of $w$ and $\zeta$, the interior penalty method is applied along the element boundaries \cite{Brenner,Engel}. We used parameter continuation in $\varepsilon$ for domains with different $W$ values. We call the planar configuration of the sheet, i.e. $w\equiv 0$, the \emph{trivial} solution. During continuation the smallest eigenvalue of the numerically computed Jacobian of $F(\hat{\xu}_\Delta,w_\Delta)$ is monitored. A positive-definite Jacobian indicates stable configuration. Hence, the point of disappearance of wrinkles was determined by seeking numerically the value $\varepsilon=\varepsilon_2$ at which the trivial solution regains stability (i.e. the smallest eigenvalue becomes positive) upon loading or loses stability upon unloading. 

\begin{figure}
    \centering
    \begin{subfigure}[b]{0.49\textwidth}
			\includegraphics[width=0.99\textwidth]{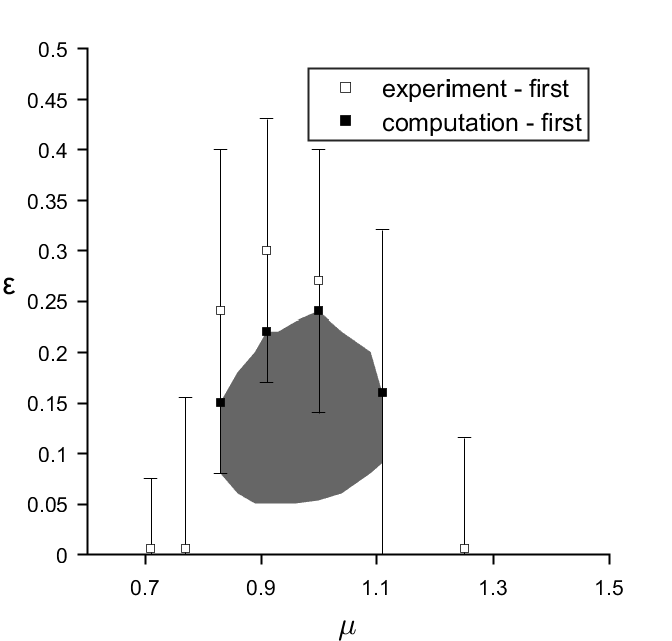}
			\caption{first loading}
    \end{subfigure}
    \begin{subfigure}[b]{0.49\textwidth}
			\includegraphics[width=0.99\textwidth]{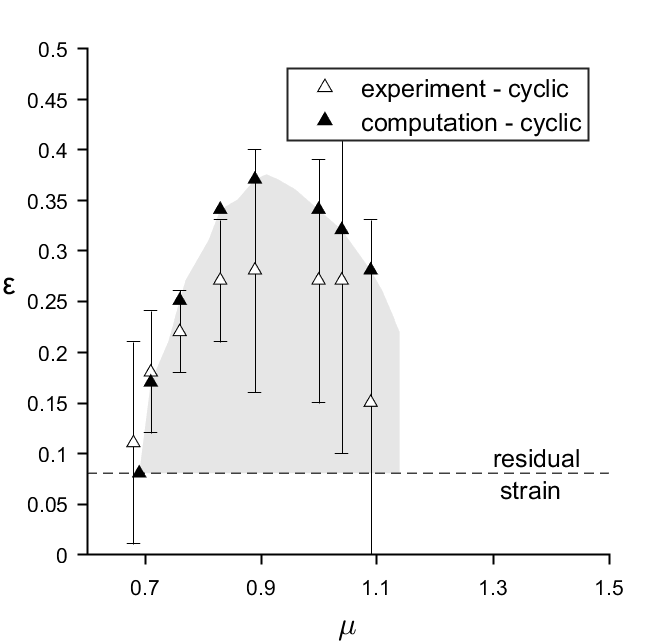}
			\caption{unloading and consecutive loading cycles}
    \end{subfigure}
    \caption{Comparison of experimental and numerical results: points of disappearance of wrinkles on unloading, $\varepsilon$ vs. $\mu$ with error bars.}\label{fig:result1}
\end{figure}

\begin{figure}[!ht]
\begin{center}
			\includegraphics[width=0.70\textwidth]{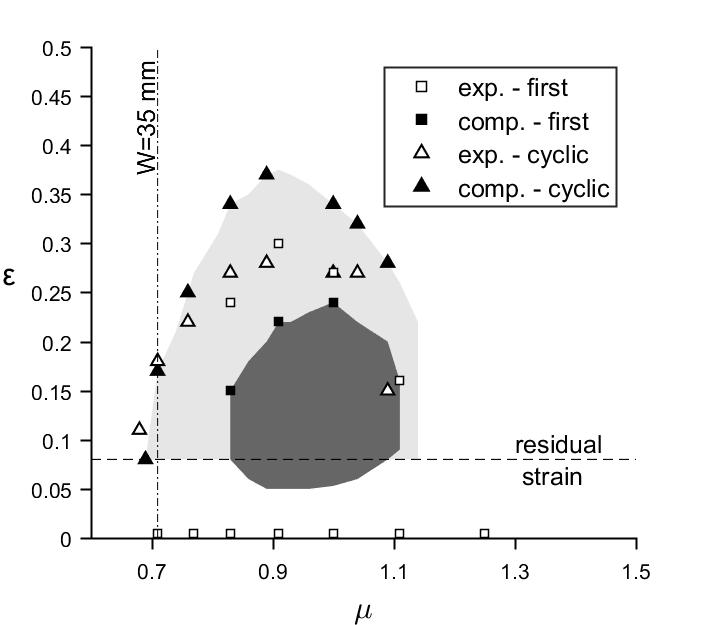}
			\caption{Both in experiments and numerical computations around $\mu\cong0.7$ wrinkling appears only during cyclic loading.} \label{fig:result2}
\end{center}
\end{figure}

Numerical results for the critical stretch values $\varepsilon_2$ are plotted against experimental data in Figure \ref{fig:result1} for both the first and cyclic loading, respectively. Note that at $\mu\cong 0.70$ ($W=35$mm), our model predicts no wrinkles during the first loading while wrinkles appear upon unloading. This is further illustrated in Figure \ref{fig:result2}, where the mean of the measured values of $\varepsilon_2$ for the first and cyclic loading are plotted against the predictions of the new model.

\begin{figure}
    \centering
    \begin{subfigure}[b]{0.49\textwidth}
			\includegraphics[width=0.99\textwidth]{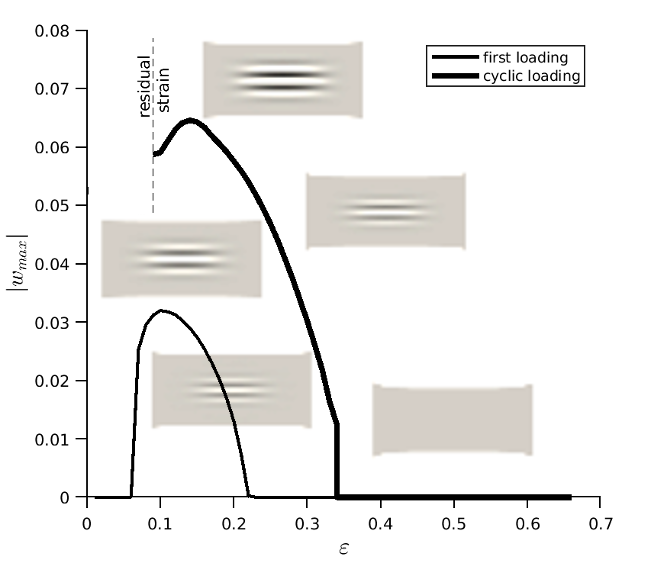}
			\caption{$\mu=1.00$}
    \end{subfigure}
    \begin{subfigure}[b]{0.49\textwidth}
			\includegraphics[width=0.99\textwidth]{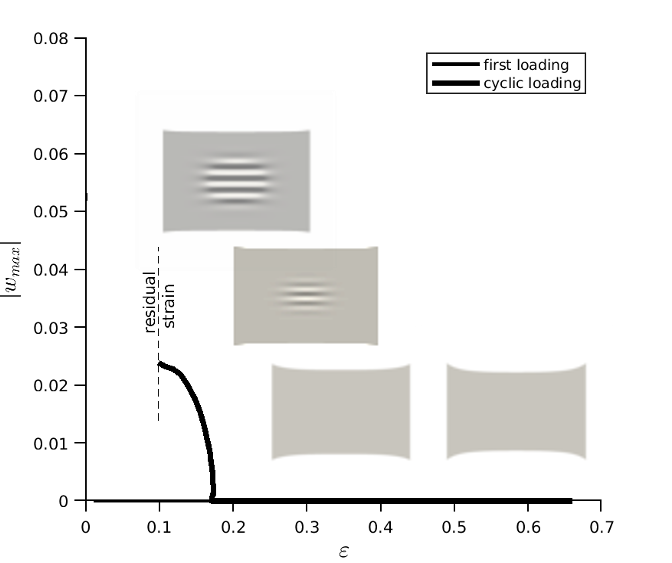}
			\caption{$\mu=0.70$}
    \end{subfigure}
    \caption{Maximum wrinkling amplitude vs. macroscopic strain, $|w_{\max}|$ vs. $\varepsilon$, for sheets with $\mu=1.0$ and $\mu=0.70$. For $\mu=0.70$ there is no wrinkling during the first loading. }\label{fig:ampl}
\end{figure}

As an illustration of our numerical results, we provide two global bifurcation diagrams in Figure \ref{fig:ampl} for $W=25$mm ($\mu=1.00$) and $W=35$mm ($\mu=0.70$) along with accompanying wrinkled configurations. In particular, we note that our model correctly recovers the observed phenomenon of Figure \ref{fig:motiv}: In both cases the sheet is still wrinkled at the residual-strain value, below which our model is no longer valid – the sheet becomes slack. We also note that for $\mu=0.70$ there is no wrinkling during the first loading, while wrinkles emerge upon unloading. 

Finally, we demonstrate that material orthotropy emerges in our model. Following the notation of \cite{Sipos}, let $r$ denote the \emph{degree of orthotropy}, defined by
\begin{equation}
\label{eq:def_r}
r:=\frac{E_{90}}{E_0},
\end{equation}

\noindent where $E_0$ and $E_{90}$ denote the tangential moduli of elasticity in directions $x$ and $y$, respectively. In the experiments we found that after unloading from $\varepsilon_{\max}=0.66$, the degree of orthotropy reached $r\cong 1.80$ \cite{Sipos}. Note that $r$ was measured after cyclic loading of the specimen; it was completely unloaded and a pull test was carried out either in direction $x$ or $y$. The degree of orthotropy $r$ was computed from the tangential moduli of elasticity measured at the initiation of loading. (Some specimens were tested in the $45^\circ$ direction to obtain the shear modulus, cf. \cite{Sipos}.) 

The pull test in the $y$ direction in our model can be associated with $C_{11}=\lambda_2^{-1}$ and $C_{22}=\lambda_2^2$, where $\lambda_2$ denotes the stretch applied in the $x$ direction (on the formally stretched specimen in the $x$ direction). The engineering stress $T_{90}$ in the $y$ direction can be expressed as

\begin{equation}
\label{eq:T90}
T_{90}(\lambda_2)=2\,{\frac {\alpha{\lambda_2}^{4}+\beta\eta\lambda_2^{3}-\alpha\lambda_2-\beta\eta}{\lambda_2^{3}}}.
\end{equation}

\noindent From (\ref{eq:T0}) and (\ref{eq:T90}) the moduli of elasticity are found to be 

\begin{align}
\label{eq:Y0Y90}
E_0(\lambda_1)&=\frac{\partial T_0(\lambda_1)}{\partial\lambda_1}=\frac{2\alpha(d+\eta-d\eta)\lambda_1^4+2\alpha(1+\eta)\lambda_1+6\beta\eta}{\lambda_1^4},\\
E_{90}(\lambda_2)&=\frac{\partial T_{90}(\lambda_2)}{\partial\lambda_2}=\frac{2\alpha\lambda_2^4\eta+4\alpha\lambda_2+6\beta\eta}{\lambda_2^4}. 
\end{align}

\noindent Substituting our model parameters at $\eta=1$ yields $r=1$ (regardless of the values of $\lambda_1$ and $\lambda_2$), i.e., the model exhibits isotropy on the unloaded specimen during the first loading, as expected. To predict the degree of orthotropy after the cyclic loading one needs to compensate the residual strain. We used $\lambda_1=1.08$ and $\lambda_2=1.08^{-1/2}$ in our calculations. Recall that unloading after the loading cycles results in $\eta=0.88$. With these in hand, our model predicts $r=1.40$ for the degree of orthotropy, which fairly agrees with the measurements, as expected.

\section{Conclusion}
A finite-deformation pseudo-elastic model accounting for the Mullins effect is introduced to qualitatively and quantitatively explain experimental data measured on highly stretched, thin elastomer sheets. Recognizing the anisotropic nature of damage, a simple model, characterized by a single state variable, is tuned to capture the measured residual strain and stress softening behavior, as well as the measured re-emergent wrinkling behavior observed upon unloading. Our motivating experimental observation, namely, that the first appearance of wrinkles for certain aspect ratios occurs during the first unloading of the specimen, is then accurately predicted by the new pseudo-elastic model. By considering the dominance of the stress in the main stretch direction, the classical pseudo-elastic model with two dissipation fields can be significantly simplified such that the model accurately predicts the residual strain/stress softening and wrinkling behavior recorded in the measurements.

\section*{\small Acknowledgments}
The work of A.A.S was supported by the J\'anos Bolyai Research Scholarship of the Hungarian Academy of Sciences. The work of T.J.H. was supported in part by the National Science Foundation through grant DMS-1613753. We thank \emph{K\'aroly P. Juh\'asz} and \emph{Ott\'o Sebesty\'en} for their valuable help in the experimental work. The Zwick Z150 material testing machine was provided by the T\'AMOP 4.2.1/B-09/1/KMR-2010-0002 grant. 

\section*{\small Author contributions}
T.J.H. initiated the research, E.F. carried out the experiments, T.J.H. and A.\'A.S. developed the model. E.F., T.J.H. and A.\'A.S. wrote the paper.


\appendix
\section{Connection with the general model of (Dorfmann and Ogden, 2004)}
A pseudoelastic model of an incompressible medium exhibiting stress-softening and residual strain is presented in \cite{Dorfman}. Equation (41) there reads
\begin{equation}
\label{eq:dorman_model}
W(\lambda_1,\lambda_2,\eta_1,\eta_2)=\eta_1W_0(\lambda_1,\lambda_2)+(1-\eta_2)N(\lambda_1,\lambda_2)+\hat{\Phi}(\eta_1,\eta_2),
\end{equation}

\noindent where $W_0(\lambda_1,\lambda_2)$ is the energy density function of a 2D incompressible hyperelastic material, $\lambda_1$, $\lambda_2$ are the proncipal stretches,
the function $N$ is introduced in order to exhibit residual strain and $\hat{\Phi}(\eta_1,\eta_2)$ is the dissipation function. It is postulated, that the state variables $\eta_1$ and $\eta_2$ are unity as long as unloading has not been started and in that case the model recovers elasticity as $W(\lambda_1,\lambda_2,1,1)=W_0(\lambda_1,\lambda_2)$ with $\hat{\Phi}(1,1)=0$. Increasing damage is associated with decrease of the dissipation fields $\eta_1$ and $\eta_2$. Our model in equation (\ref{model:mED}) belongs to this model family. To see this, first let us reduce the number of dissipation fields via 
\begin{equation}
\label{eq:dorman_model_eta2}
\eta_2:=(1+q)\eta_1-q,
\end{equation}

\noindent where $q$ is some fixed parameter. Obviously $\eta_1=1 \Leftrightarrow \eta_2=1$ holds. Now $\hat{\Phi}(\eta_1,\eta_2)$ is simplified to $\Phi(\eta_1)$. Let $W_0$ stand for an isotropic, incompressible Mooney-Rivlin material with material parameters $\alpha$ and $\beta$. Finally, let $N$ be an incompressible, modified Mooney-Rivlin material with material parameters $\alpha,\beta,v_1,v_2,v_3$, viz.,
\begin{eqnarray}
\label{eq:dorman_model_MR}
W_0:=\alpha\left[\tr\xC+\dfrac{1}{\det{\xC}}-3\right]+\beta\left[\dfrac{\tr\xC}{\det{\xC}}+\det{\xC}-3\right],\\
N:=\alpha\left[v_1(C_{11}-1)+v_2(C_{22}-1)+v_3\left(\dfrac{1}{\det\xC}-1\right)\right]+\beta v_2\left[\dfrac{\tr\xC}{\det{\xC}}+\det{\xC}-3\right].
\end{eqnarray}

\noindent Let $v_1=1$, $v_2\neq 1$, and $v_3=(1+q)^{-1}$ and then set
\begin{eqnarray}
q:=\frac{1+d-v_2-dv_2}{v_2+dv_2-1},\\
\eta_1:=\eta-\frac{dv_2(1-\eta)}{v_2-1},
\end{eqnarray}

\noindent where $d$ and $\eta$ are the parameter and the state variable, respectively, appearing in (\ref{model:mED}). Now equation (\ref{eq:dorman_model_eta2}) can be written as:
\begin{equation}
\eta_2:=1-d\frac{1-\eta}{v_2-1}.
\end{equation}

Note that $\eta=1 \Leftrightarrow \eta_1=1$ and $\eta=1 \Leftrightarrow\eta_2=1$ hold. Substituting the formulas for $\eta_1$, $\eta_2$, $W_0$ and $N$ into (\ref{eq:dorman_model}) yields the membrane energy (\ref{model:mED}) plus the dissipation term $\Phi(\eta)$ as in (\ref{model:ED}).


\end{document}